\documentclass[a4paper,11pt]{article} 

\usepackage{jinstpub} 
\pdfoutput=1 
\title{\boldmath Subpixel Response of SOI Pixel Sensor for X-ray Astronomy with Pinned Depleted Diode: First Result from Mesh Experiment}

\author[a,1]{K.~Kayama,\note{Corresponding author.}}
\author[a]{T.~G.~Tsuru,}
\author[a]{T.~Tanaka,}
\author[a]{H.~Uchida,}
\author[a]{S.~Harada,}
\author[a]{T.~Okuno,}
\author[a]{Y.~Amano,}
\author[b]{J.~S.~Hiraga,} 
\author[b]{M.~Yoshida,}
\author[b]{Y.~Kamata,}
\author[b]{S.~Sakuma,}
\author[b]{D.~Yuhi,}
\author[b]{Y.~Urabe,}
\author[c]{H.~Tsunemi,}
\author[d]{H.~Matsumura,}
\author[e]{S.~Kawahito,}
\author[e]{K.~Kagawa,}
\author[e]{K. ~Yasutomi,}
\author[e]{S.~Shrestha,}
\author[e]{S.~Nakanishi,}
\author[f]{H.~Kamehama,}
\author[g]{Y.~Arai,}
\author[g]{I.~Kurachi,}
\author[h]{A.~Takeda,}
\author[h]{K.~Mori,}
\author[h]{Y.~Nishioka,}
\author[h]{K.~Fukuda,}
\author[h]{T.~Hida,}
\author[h]{M.~Yukumoto,}
\author[i]{T.~Kohmura,}
\author[i]{K.~Hagino,}
\author[i]{K.~Oono,}
\author[i]{K.~Negishi,}
\author[i]{K.~Yarita,}
\affiliation[a]{Department of Physics, Graduate School of Science, Kyoto University,\\Kitashirakawa, Sakyo-ku, Kyoto, 606-8502, Japan}
\affiliation[b]{Department of Physics, School of Science and Technology, Kwansei Gakuin University,\\2-1 Gakuen, Sanda, Hyogo, 669-1337, Japan}
\affiliation[c]{Department of Earth and Space Science, Graduate school of Science, Osaka University,\\1-1 Machikaneyama-cho, Toyonaka-shi, Osaka 560-0043, Japan}
\affiliation[d]{Kavli IPMU, The University of Tokyo,\\5-1-5 Kashiwanoha, Kashiwa, Chiba, 277-8583, Japan}
\affiliation[e]{Research Institute of Electronics, Shizuoka University,\\3-5-1 Johoku, Naka-ku, Hamamatsu 432-8011, Japan}
\affiliation[f]{Information and Communication Systems Engineering, Okinawa National College of Technology,\\905 Henoko, Nago, Okinawa, Japan}
\affiliation[g]{Institute of Particle and Nuclear Studies, High Energy Accelerator Research Org., KEK,\\1-1 Oho, Tsukuba 305-0801, Japan}
\affiliation[h]{Department of Applied Physics and Electronic Engineering, Faculty of Engineering, University of Miyazaki,\\1-1 Gakuen Kibanadai-Nishi, Miyazaki, 889-2192, Japan}
\affiliation[i]{Department of Physics, Faculty of Science and Technology, Tokyo University of Science,\\2641 Yamazaki, Noda, Chiba 278-8510, Japan}
\emailAdd{kayama.kazuho.57r@kyoto-u.jp}
\abstract{
We have been developing a monolithic active pixel sensor, ``XRPIX'', for the Japan led future X-ray astronomy mission ``FORCE'' observing the X-ray sky in the energy band of 1--80~keV with angular resolution of better than $15''$. 
XRPIX is an upper part of a stack of two sensors of an imager system onboard FORCE, and covers the X-ray energy band lower than 20~keV. 
The XRPIX device consists of a fully depleted high-resistivity silicon sensor layer for X-ray detection, a low resistivity silicon layer for CMOS readout circuit, and a buried oxide layer in between, which is fabricated with ${\rm 0.2~\mu m}$ CMOS silicon-on-insulator (SOI) technology. 
Each pixel has a trigger circuit with which we can achieve a ${\rm 10~\mu s}$ time resolution, a few orders of magnitude higher than that with X-ray astronomy CCDs. 
We recently introduced a new type of a device structure, a pinned depleted diode (PDD), in the XRPIX device, and succeeded in improving the spectral performance, especially in a readout mode using the trigger function. 
In this paper, we apply a mesh experiment to the XRPIX devices for the first time in order to study the spectral response of the PDD device at the subpixel resolution. 
We confirmed that the PDD structure solves the significant degradation of the charge collection efficiency at the pixel boundaries and in the region under the pixel circuits, which is found in the single SOI structure, the conventional type of the device structure. 
On the other hand, the spectral line profiles are skewed with low energy tails and the line peaks slightly shift near the pixel boundaries, which contribute to a degradation of the energy resolution. 

 }
\keywords{X-ray detectors, Space instrumentation, Imaging spectroscopy}
\proceeding{The 9$^{\text{th}}$  International Workshop on Semiconductor Pixel Detectors for Particles and Imaging (PIXEL 2018)\\
  December 10 - 14, 2018 \\
  Academia Sinica, Taipei, Taiwan}
\begin{document}
\maketitle
\flushbottom

\section{Introduction}
We have been developing a new sensor for the future X-ray astronomy satellite mission led by Japan, Focusing On Relativistic universe and Cosmic Evolution (FORCE)~\cite{mori2016, nakazawa2018}.
The primary scientific objective is to trace the cosmic formation history by searching for ``missing black holes'' in various mass-scales including ``buried supermassive black holes (SMBHs)'', ``intermediate-mass black holes'', and ``isolated stellar-mass black holes'' in our Galaxy. 
Broadband observations are essential to uncover the buried SMBHs, to measure the masses of the black holes, and to distinguish isolated stellar-mass black holes from neutron stars and white dwarfs. 
Thus, the FORCE mission is designed to perform broadband X-ray imaging spectroscopy from $\rm 1~keV$ to $\rm 80~keV$. 
The FORCE mission carries high-resolution hard X-ray mirrors and wide-band hybrid X-ray imagers. 
The imager system consists of a stack of silicon and CdTe sensors, which detect X-rays with the energies lower and higher than $\rm 20~keV$, respectively. 

We have been developing the silicon sensor for the FORCE mission. 
In order to reduce non-X-ray backgrounds (NXB), we adopt the anti-coincidence technique in which we use the signals from scintillator enclosing the imager as veto. 
Since the counting rate of the scintillator is estimated to be $\sim 10$~kHz in orbit, the sensors are required to have a ${\rm \sim 10}$-${\mu s}$ time resolution, which is impossible with charge-coupled devices (CCDs), the standard silicon sensors used in X-ray astronomy~\cite{struder2001, garmire2003, koyama2007, tanaka2018}. 
We, therefore, have been developing a new type of active pixel sensors, called ``XRPIX''~\cite{tsuru2018}, fabricated using the ${\rm 0.2\ \mu m}$ CMOS Silicon-On-Insulator (SOI) technology~\cite{arai2011}.
The XRPIX device consists of three layers: a fully depleted high-resistivity silicon sensor layer for X-ray detection, a low resistivity silicon layer for CMOS readout circuit, and a buried oxide (BOX) layer in between.
Each pixel has a self-trigger output function, which allows us to detect X-rays at a time resolution better than $10$~$\mu{\rm s}$, and to read out only triggering pixels.
We refer to the readout mode using the trigger function as the ``Event-Driven readout mode''. 

Among the performances required for FORCE, the depletion thickness of the sensor layer and the Event-Driven readout function itself were realized in the early stages of the development. 
On the other hand, the spectral performance has not met the requirement of FORCE, especially in the Event-Driven readout mode. 
It is because the ``Single SOI'' (SSOI) structure, the conventional type of the device structure, has difficulties in reducing the sensing-node capacitance and the interference between the circuit layer and the sensor layer. 
In order to solve this problem, we developed two new types of device structure having an electrical shielding layer between the sensor layer and the circuit layer: one is a ``Double SOI'' (DSOI) structure~\cite{lu2016, miyoshi2017, miyoshi2018} and the other is a ``Pinned Depleted Diode'' (PDD) structure developed by Kamehama et~al. (2018)~\cite{kamehama2018}. 
Applying the structures to the XRPIX devices, we found that both devices improved the spectral performance in the Event-Driven readout mode~\cite{tsuru2018, harada2019, takeda2019}. 


In this paper, we examine the response of XRPIX6E at the subpixel level~\cite{harada2019}, the first XRPIX device having the PDD structure. 
While we have used synchrotron radiation facilities for subpixel studies~\cite{matsumura2015, negishi2019, hagino2019a, hagino2019b}, we apply so-called ``mesh experiments'' invented by Tsunemi et~al. (1997) to the XRPIX series for the first time~\cite{tsunemi1997}. 

%

\section{Device Description}\label{Device Description}
\begin{figure}[htbp]
  \centering 
  \includegraphics[width=.7\textwidth]{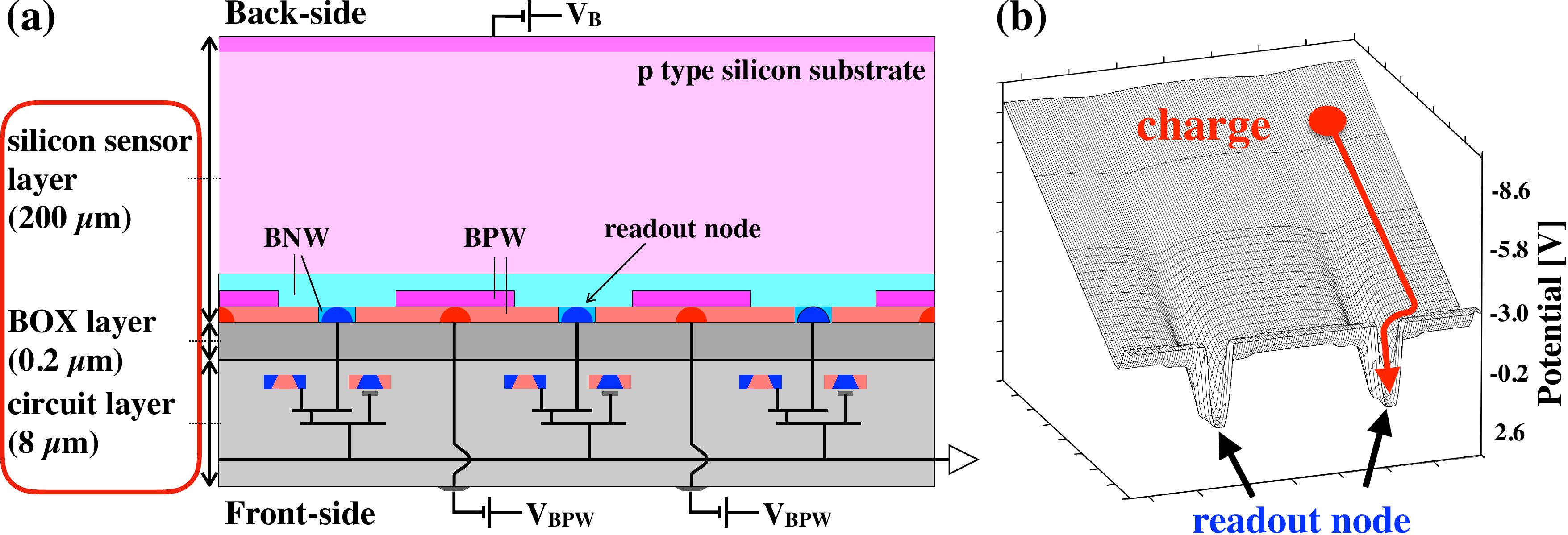}
  \caption{(a) Schematic cross-sectional view of XRPIX6E. This figure is adopted from Harada et~al. (2019)~\cite{harada2019}. (b) Simulation of the potential in the sensor layer.}
	\vspace*{-0.5\intextsep} 
  \label{fig:1_2_1}
\end{figure}

XRPIX6E is the first model of XRPIXs equipped with the PDD structure~\cite{kamehama2018, harada2019}. 
Figure~\ref{fig:1_2_1}(a) shows a schematic cross-sectional view of XRPIX6E.
The highly doped buried p-well (BPW) acts as an electrostatic shield to reduce the capacitive coupling between the charge sensing-nodes and the circuit layer. 
This structure is also effective for high charge collection efficiency (CCE).
As shown in the simulations (Figure~\ref{fig:1_2_1}b), the stepped buried n-well (BNW) help signal charge be transported to the charge sensing-node without touching the interface between the sensor layer and BOX layer. 

As described in Harada et~al. (2019)~\cite{harada2019}, XRPIX6E has the sensor layer with a thickness of 200~$\rm \mu m$, which is fabricated using p-type floating zone wafers with a resistivity of ${\rm >25~k\Omega~cm}$, and $48\times 48$~pixels with a pixel size of 36~${\rm \mu m~\times}$~36~${\rm \mu m}$ in the imaging area. 
Each pixel has analog and trigger circuits. 
The signal charge collected at the charge sensing-node is amplified by a charge sensitive amplifier in the pixel. 
A trigger signal and its column and row address are output if the pulse height exceeds the event threshold. 
The pulse heights of the pixels are read out through peripheral readout circuits consisting of a column programmable gain amplifier and an output buffer.
A detailed explanation of the pixel and peripheral readout circuits is described in Takeda et~al. (2019)~\cite{takeda2019}. 

\section{Experiments and Results}
\subsection{X-ray spectra of $^{57}$Co}\label{Co-57spec} 
\begin{figure}[bhtp]
  \centering 
	  \includegraphics[width=.7\textwidth]{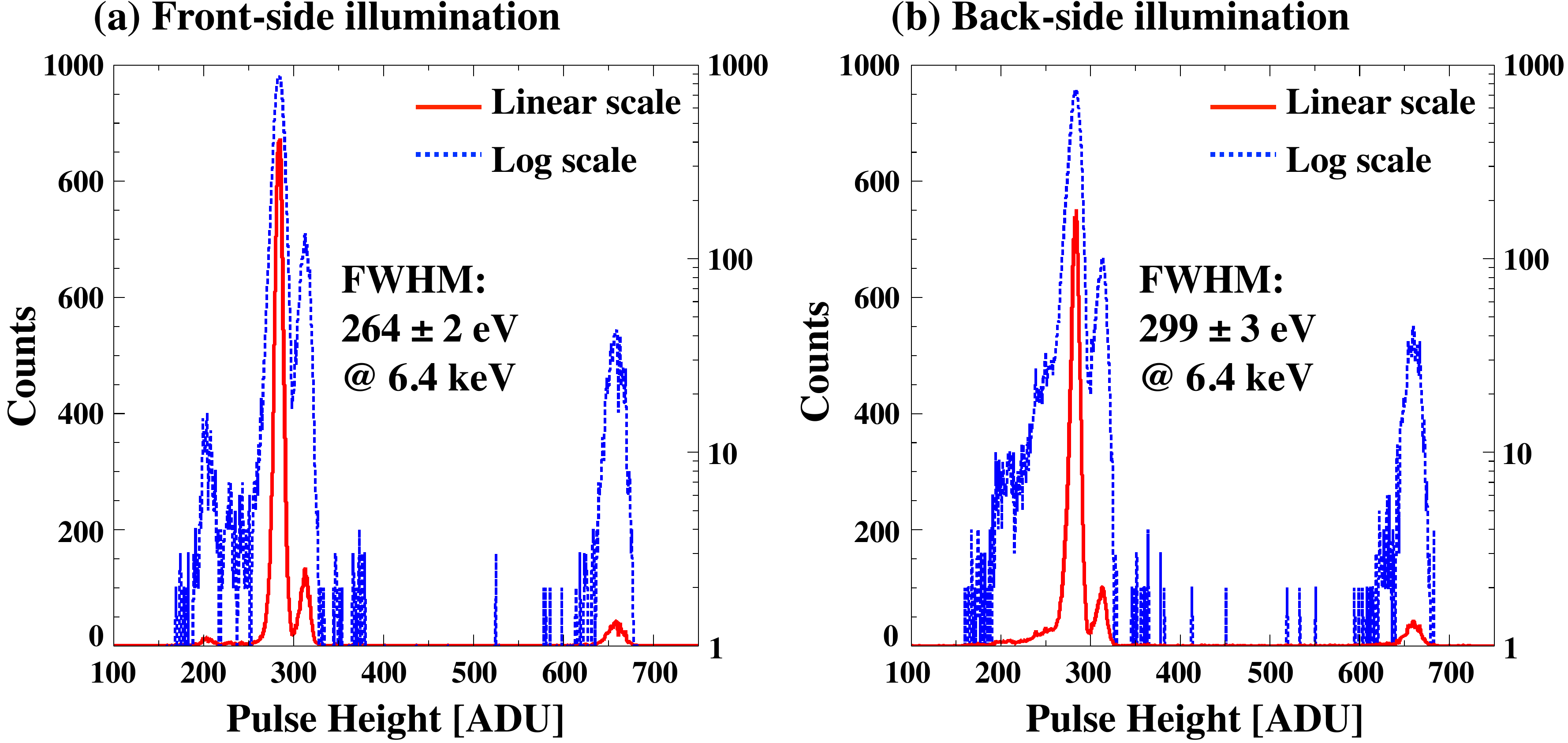}
	\vspace*{-0.5\intextsep} 
  \caption{\label{fig:1_2_2}$^{57}$Co X-ray spectra of single-pixel events with FI (left) and BI (right).}
\end{figure}

Harada et~al. (2019) reported the first results from the evaluation of XRPIX6E~\cite{harada2019}.  
After their paper, we have further optimized the operation of the device by adjusting voltages applied to BPW. 
Figure~\ref{fig:1_2_2} shows optimized single-pixel event spectra of $^{57}$Co X-rays. 
The device was cooled to $-60$~\textcelsius\ with a thermostatic chamber to reduce the dark current. 
We applied a back-bias voltage of $\rm V_b = -200~V$ to fully deplete the sensor layer. 
We operated the device in the Event-Driven readout mode following the sequence described in Takeda et~al. (2013)~\cite{takeda2013}. 
We read out the pulse heights of the triggering pixel and the $8\times 8$ pixels having the triggering pixel at their centers as an X-ray event. 
We collected X-ray events from the entire imaging area of $48\times 48$ pixels, while only $8\times 8$ pixels located at the center of the device were employed in Harada et~al. (2019). 
We analyzed the X-ray event data using the methods described by Ryu et~al. (2011) and Nakashima et~al. (2012)~\cite{ryu2011, nakashima2012}. 
Since signal charge diffuses during the drift in the depletion region, the signal charge sometimes spreads into neighboring pixels; this is called a split event. 
Therefore, we classified each X-ray even into one of the following types according to patterns of the split: ``single-pixel'' (no split), ``vertically split 2-pixel event'', ``horizontally split 2-pixel event'', ``3- or 4-pixel event'' and ``others''. 
The split-threshold was set to 10~ADU, which is equivalent to 0.2~keV. 


The energy resolutions of single-pixel events at 6.4~keV are 264$\rm~\pm~2~eV$ and 299$\rm~\pm$~3~eV in FWHM with a front-side illumination (FI) and with a back-side illumination (BI), respectively. 
They are significantly better than those in Harada et al. (2019)~\cite{harada2019} thanks to the optimization. 
We note that the energy resolution with BI meets the FORCE requirement of 300~eV in FWHM at 6~keV~\cite{nakazawa2018}. 

\subsection{Mesh Experiment}
\begin{wrapfigure}{r}{.40\textwidth}
  \centering 
  \includegraphics[width=.38\textwidth]{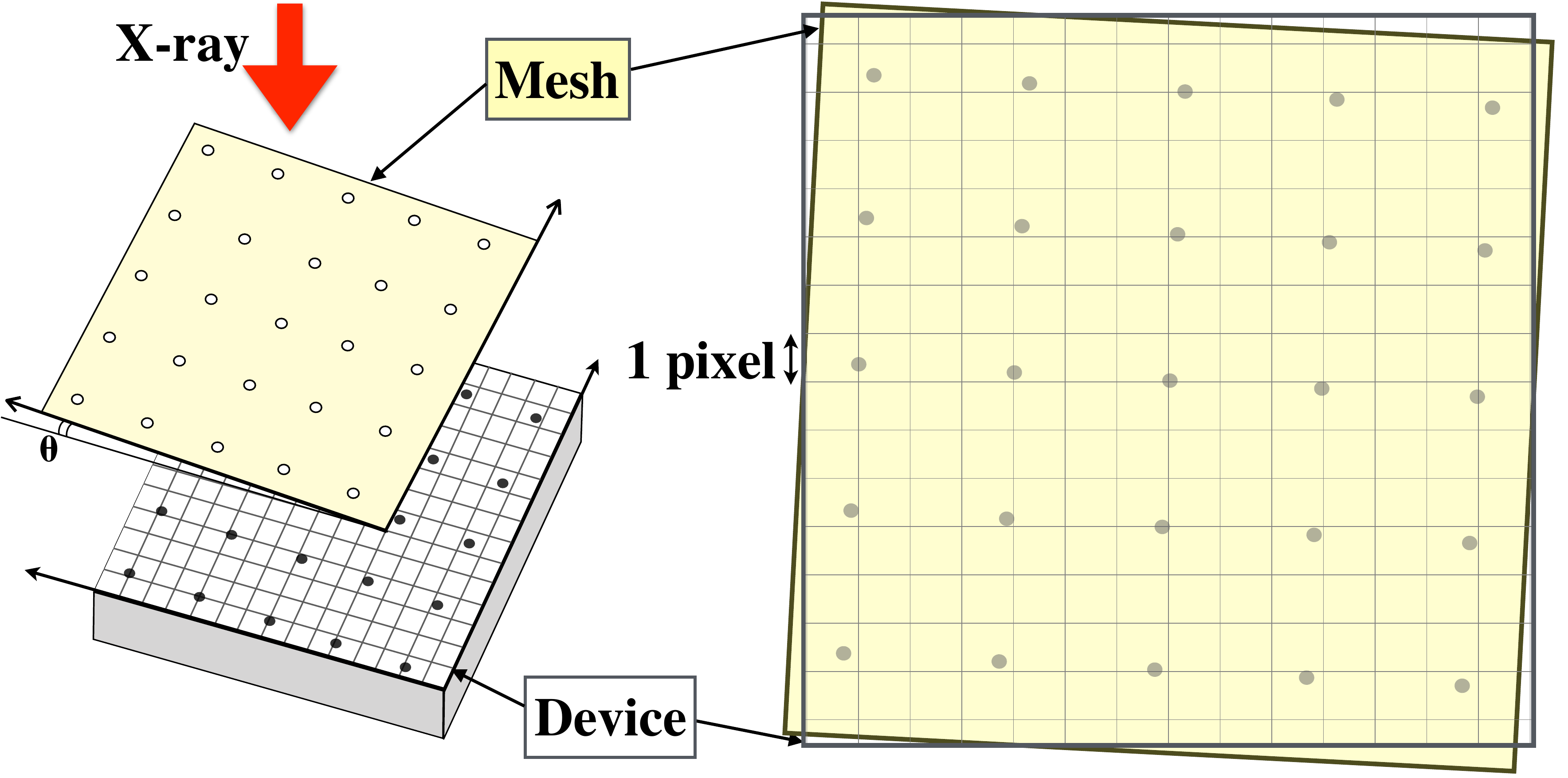}
\vspace*{-\intextsep} 
  \caption{\label{fig:2_1_1} Schematic view of the mesh experiment. This figure is adopted and modified from Tsunemi~et~al. (1998)~\cite{tsunemi1998}}
\end{wrapfigure}

The principle of the mesh experiment we applied is introduced in Tsunemi et~al. (1997) and Tsunemi et~al. (1998)~\cite{tsunemi1997, tsunemi1998}. 
Figure~\ref{fig:2_1_1} shows a schematic view of the mesh experiment. 
We place a metal mesh having evenly spaced holes over the XRPIX device and expose it to a quasiparallel X-ray beam. 
The holes limit the X-ray exposure to specified subpixel locations in a pixel. 
Rotating the mesh by a small angle with respect to the device produces a moire pattern of the X-ray landing location in the imaging area of the device. 
Referring to the moire pattern in an X-ray image taken with the device, we can determine the mutual alignment between the mesh and the device, the specified subpixel location of each hole, and thus the subpixel landing position of each X-ray event. 
By collecting the X-ray events obtained from the entire imaging area and stacking them into a ``representative pixel'', we can measure response of the device at the subpixel resolution. 

\begin{figure}[bhtp]
  \centering 
  \includegraphics[width=.75\textwidth]{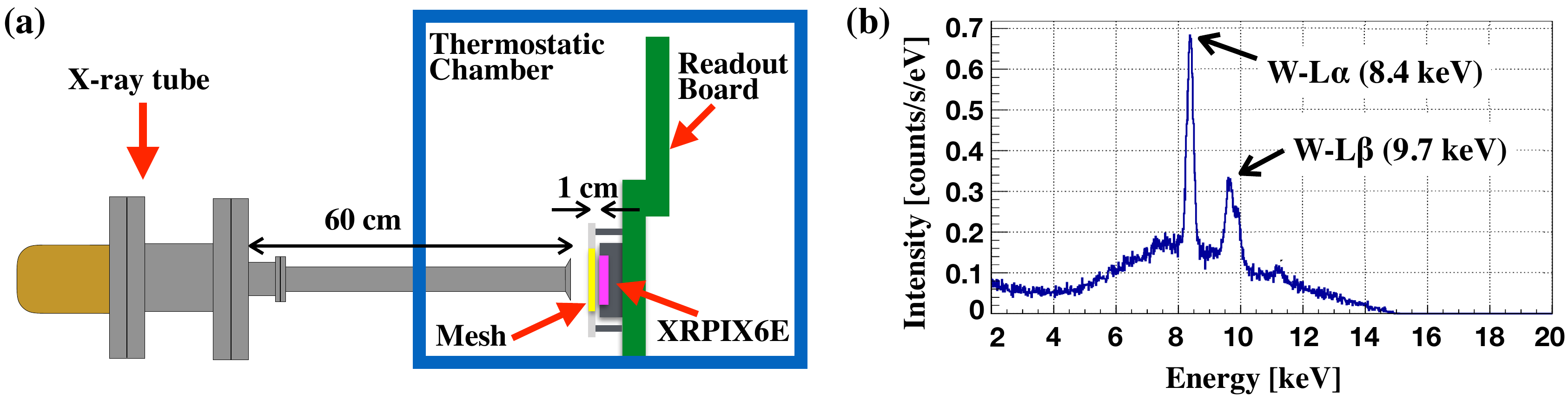}
	\vspace*{-\intextsep} 
  \caption{\label{fig:2_2_1} (a) Schematic view of the experimental setup. (b) X-ray spectrum with a silicon drift detector.}
\end{figure}

In this paper, we placed the microfocus X-ray tube with a tungsten anode 60~cm from the device as shown in Figure~\ref{fig:2_2_1}(a). 
Figure~\ref{fig:2_2_1}(b) shows an X-ray spectrum obtained with a silicon drift detector, where we see W--L$\alpha$ and L$\beta$ lines, and a bremsstrahlung component. 
We used the gold mesh processed by microworks GmbH with a thickness of $\rm 80~\mu m$ and placed it 1~cm from the device. 
The mesh has holes with a diameter of $\rm 4~\mu m$ at a pitch of 108~$\rm \mu m$ which is three times the pixel size of 36~$\rm \mu m$. 
We do not discuss absolute detection efficiency in this paper since the effective size of each hole has not been calibrated yet. 
Applying the same operational condition as in Section~\ref{Co-57spec}, we obtained X-ray events with FI and BI. 
We, then, determined the mesh alignment with respect to the device following the method described in Tsunemi~et~al. (1998)~\cite{tsunemi1999}.

\begin{figure}[bhtp]
  \centering 
  \includegraphics[width=.66\textwidth]{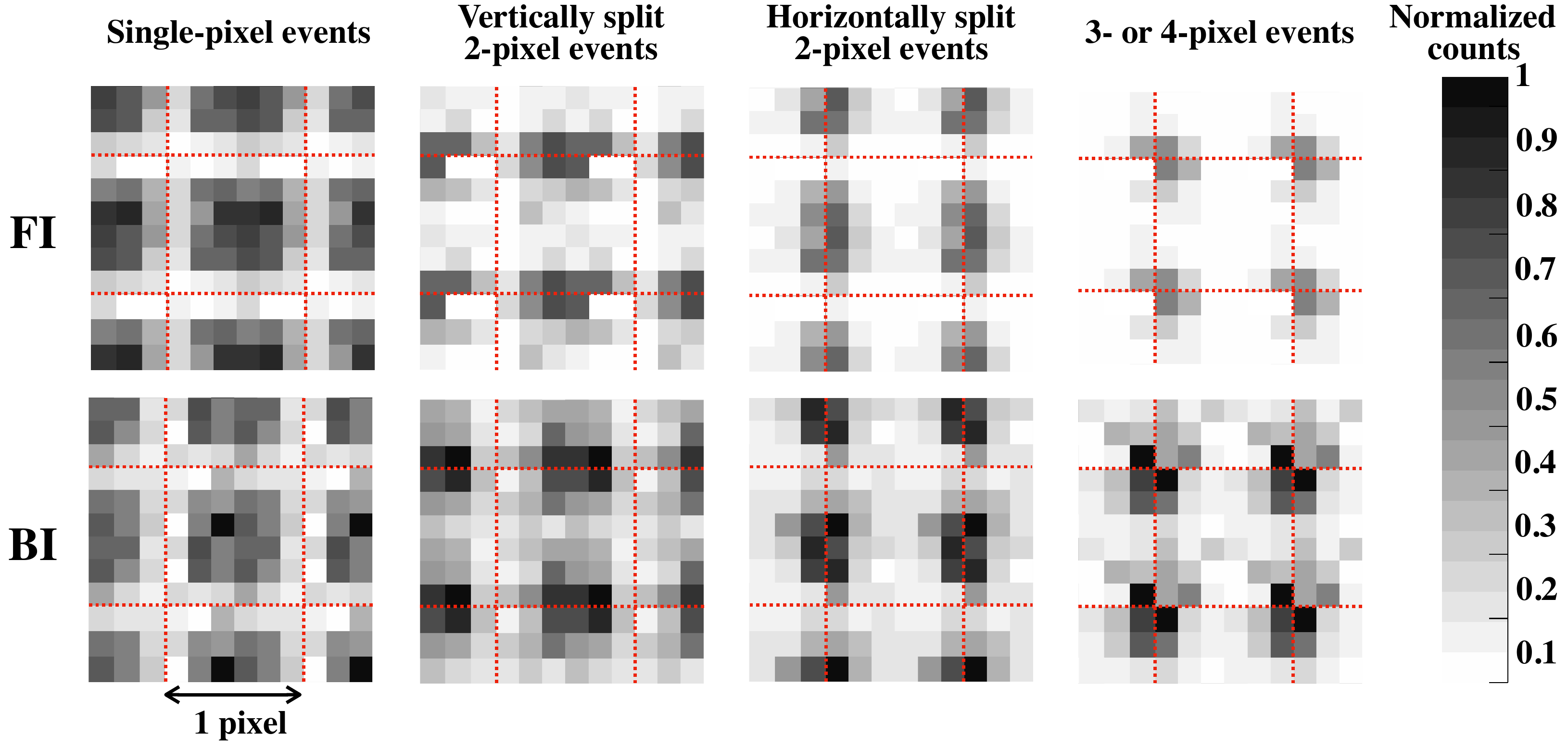}
	\vspace*{-\intextsep} 
  \caption{\label{fig:5_1_1_4} Distribution of various events in the representative pixel: the single-pixel, vertically split 2-pixel, horizontally split 2-pixel, and 3- or 4-pixel split events (columns from left to right) and those with FI and BI (top and bottom rows).}
\end{figure}


Figure~\ref{fig:5_1_1_4} shows the distribution of various events in the representative pixel that is divided into $6\times 6$ subpixel regions with a pitch of ${\rm 6~\mu m}$.  
It covers the $2\times 2$ representative pixel area for clarity. 
The single-pixel events are dominant in the center region of the pixel with a ratio of $0.6$--$0.8$, while the multi-pixel events are dominant near the pixel boundaries. 
This results indicate that the representative pixel is reconstructed correctly as expected. 
We also notice that the fraction of single-pixel events with FI is higher than that with BI. 
The average charge cloud of the W--L${\rm \alpha}$ X-rays at ${\rm 8.4~keV}$ with BI is larger then that with FI, since the attenuation length ${\rm \sim 80~\mu m}$ for the X-rays is significantly smaller than the thickness of the sensor layer of $200\ {\mu m}$. 
Thus, the difference between FI and BI is well explained by the effect of the charge sharing due to the charge diffusion during the drift in the depletion layer, which is commonly found in other silicon pixel sensors. 


\section{Discussion}
\subsection{Comparison with the Single SOI devices}
Our previous studies have clarified that there are the following two causes of the degradation in the spectral performances of the SSOI devices. 
One is that the CCE degrades near the pixel boundaries. 
Negishi et~al. (2019)~\cite{negishi2019} reported that the spectra have large tail structures near the pixel boundaries. 
It suggests that a significant part of the signal charge is lost there. 
The other one is that the CCE degrades in the sensor region under the pixel circuit outside the buried well~\cite{matsumura2015}. 
The pixel circuit distorts the electric fields in the sensor layer and makes local minimums in the electric potentials at the interface between the sensor and BOX layers, where a part of signal charge is trapped and lost. 

\begin{figure}[bhtp]
  \centering 
  \includegraphics[width=1\textwidth]{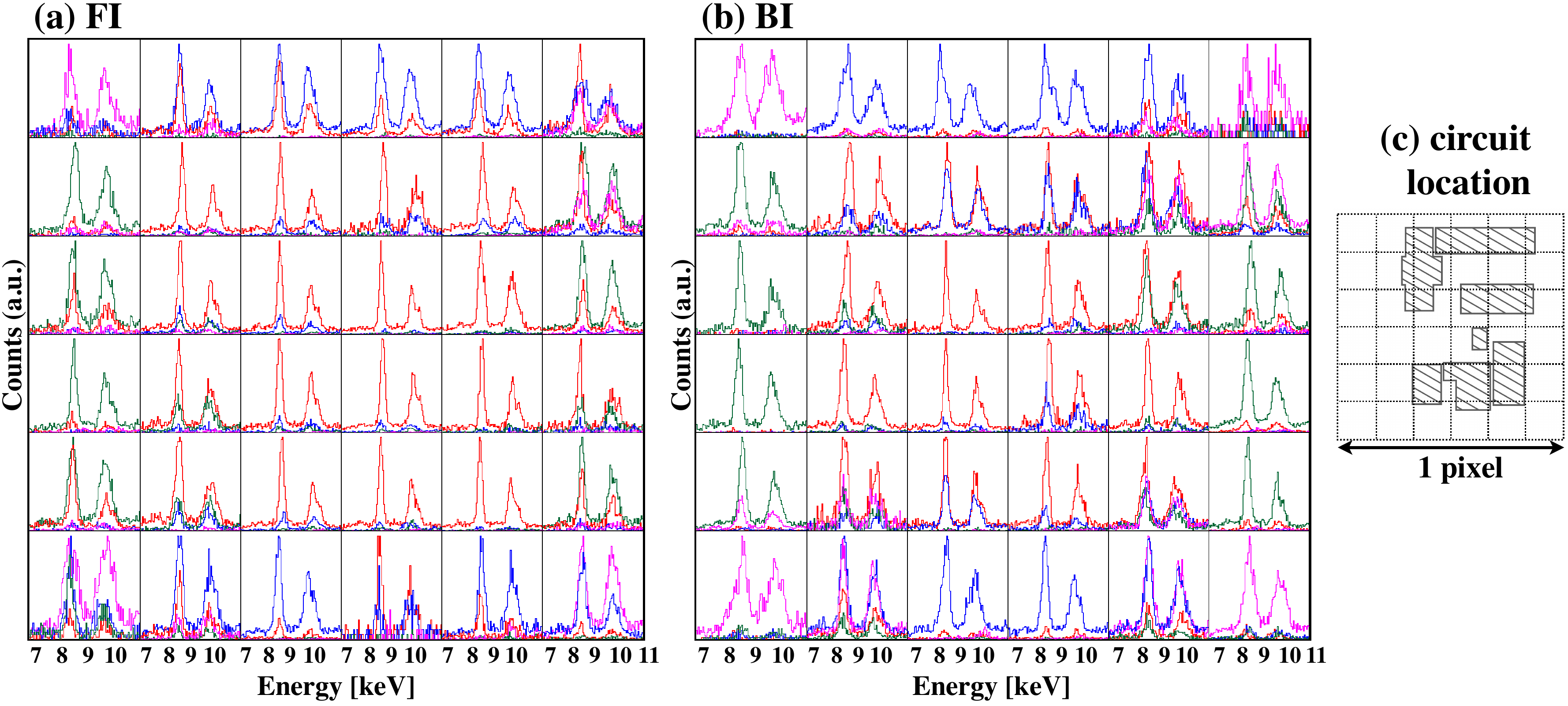}
	\vspace*{-1.5\intextsep} 
  \caption{\label{fig:4_1_1} Two-dimensional map of spectra obtained from each subpixel region with (a) the FI and (b) the BI. 
  The spectra of the single-pixel, vertically split 2-pixel, horizontally split 2-pixel and 3- or 4-pixel split events are shown in red, blue, green, magenta, respectively. 
  The counts are in arbitrary units. 
  The hatched areas in the panel (c) indicate circuit location in the pixel.
  }
\end{figure}

In order to examine the CCE of the PDD devices, we present two-dimensional maps of spectra of the single-pixel and multi-pixel events in Figure~\ref{fig:4_1_1} for $6\times 6$ subpixel regions obtained by using FI (a) and BI (b). 
In each spectrum, we show the energy range of 7--12~keV so that line shapes of W--L${\rm \alpha}$ and W--L${\rm \beta}$ can be clearly seen. 
The tail structures are much smaller than those in the SSOI devices even near the pixel boundaries. 
In Figure~\ref{fig:4_1_1}(c) we show the location of circuits in a pixel by hatched region. 
No difference is found between the spectra of the regions where the pixel circuits are present and those where the pixel circuits are not present, regardless of FI and BI. 
No large tail is seen in the spectra of the subpixel regions where the pixel circuits are located, either. 
These results imply that the stepped BNW of the PDD structure transports signal charge to the charge sensing-node almost without loss, and electric field is not distorted by pixel circuit thanks to electric shield by the BPW in the PDD structure, as designed. 

The degradation of CCE results in a low peak-to-valley ratio in the X-ray spectrum~\cite{granato2013}. 
A certain amount of background counts in channels below the peak can be observed even though the incident X-rays are monoenergetic. 
A peak-to-valley ratio is defined as the ratio of the peak to background (valley) counts, which is used as a figure of merit of a spectrometer. 
XRPIX3b, which has the best energy resolution in the SSOI devices, only exhibits a peak-to-valley ratio of $\sim 50$ with FI~\cite{takeda2015}. 
In contrast, the FI spectrum obtained with XRPIX6E exhibits a peak-to-valley ratio of $300\sim 500$ for the 6~keV X-rays as shown in Figure~\ref{fig:1_2_2}. 
We thus conclude that the large degradation of the CCE seen in the SSOI is basically improved by introducing the PDD structure. 

\subsection{Spectral Response at the Subpixel Resolution}
\begin{figure}[hbtp]
  \centering 
  \includegraphics[width=.65\textwidth]{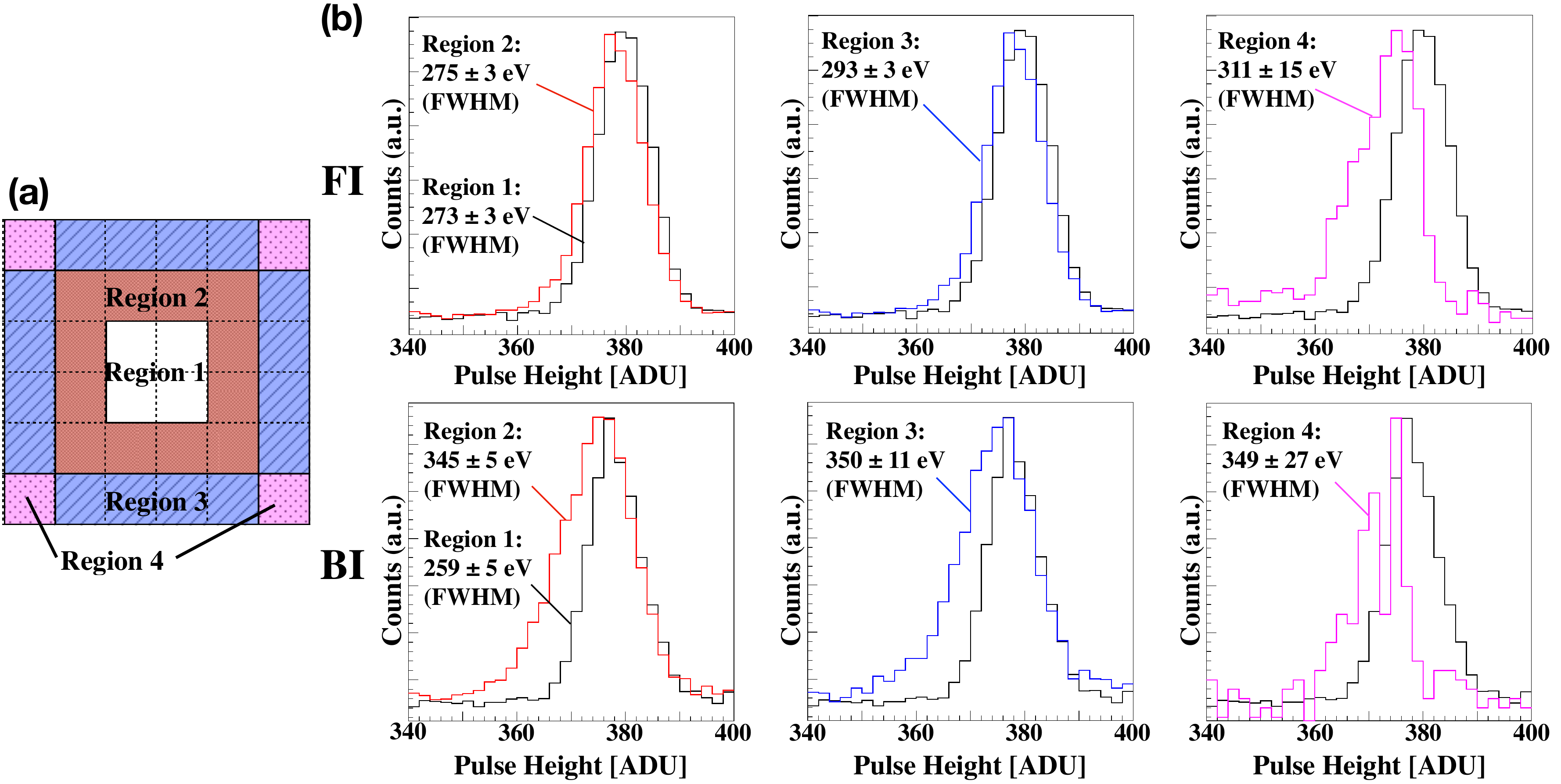}
	\vspace*{-0.5\intextsep} 
  \caption{\label{fig:5_1_1_5} 
  (a) The spectra were summed for each of the four regions shown in this panel: the pixel center (region 1), the region between the pixel center and pixel boundaries (region 2), the vertical and horizontal sides (region 3) and the pixel corners (region 4).
(b) The spectra of single-pixel events obtained at regions 2, 3 and 4 are compared to those at region 1 in the left, center and right panels, respectively. 
The upper and lower panels are spectra obtained with FI and BI, respectively.
The counts are in arbitrary units. 
The spectra at regions 1, 2, 3 and 4 are shown with the black, red, blue, and magenta lines, respectively. 
}


\end{figure}

In this section, we explore degradation of spectral performances near pixel boundaries, which is relatively smaller than those in the SSOI devices but still significantly degrades the overall spectrum. 
In order to investigate the spectral structure in detail, we divided the $6\times 6$ subpixels into four regions defined in Figure~\ref{fig:5_1_1_5}(a), and summed the spectra for each region. 

Figure~\ref{fig:5_1_1_5} shows the single-pixel event spectra around W--L${\rm \alpha}$ line for the four regions with FI and BI. 
The spectral performance of region 1 is basically the same for FI and BI. 
For FI, the degradation of energy resolution appears in region 3 and 4, while there is no difference between region 1 and 2.
For BI, significant degradation of energy resolution is clearly seen in region 2,3 and 4. 
The difference in the regions 2-4 between BI and FI would make the overall spectral performance of BI lower than FI shown in Figure~\ref{fig:1_2_2}. 

As seen in Figure~\ref{fig:5_1_1_5}, the line profiles in regions 3 and 4 with FI and regions 2--4 with BI are skewed with low energy tails, which causes the degradation of the energy resolution. 
Here, we discuss the charge sharing as a possible cause of the tail structures.
When the signal charge cloud spreads over multi pixels, the pixel with the charge amount lower than the split threshold is ignored, which results in the tail structures in the spectra. 
The tail should have a width corresponding to the split threshold at largest.
The observed tail structure indeed has a width of $\sim 10$ channels, which is consistent with the split threshold of $\sim 10$ channels which we applied. 
The effect of the charge sharing can explain the larger tail structures with BI than FI as well. 
The average charge cloud of the W--L${\rm \alpha}$ X-rays just below the boundary between the sensor layer and the BOX layer with BI is larger then that with FI since the attenuation length ${\rm \sim 80~\mu m}$ for the X-rays is significantly smaller than the thickness of the sensor layer. 
Therefore, we conclude that the charge sharing is likely the cause of the tail structures. 

In addition to the tail structures, peak shifts are seen from region 1 to region 4 for both FI and BI. 
The shift amounts to 4--6 channels corresponding to ${\rm \sim 100~eV}$ or $\sim 1$\%, which contributes enough to the degradation of the energy resolution at 8.4~keV. 
The peak shift would not be explained by the charge sharing since it is observed with FI as well as BI. 
We suppose that the charge are lost somewhere between X-ray interacting position and the charge sensing-node while being drifted along the electric field line. 
The lateral drift of the signal charge in the potential minimum shown in Figure\ref{fig:1_2_1}(b) is common to FI and BI regardless of the depth of the interacting position in the sensor layer. 
Thus, we presume that the charge loss occurs while drifting in the potential minimum region. 
In the future, we will quantitatively investigate the cause of the peak shift by changing operating conditions and X-ray energies. 
\acknowledgments
We acknowledge the manufactures of XRPIXs and valuable advice by the personnel of LAPIS Semiconductor Co., Ltd. 
This study was supported by the Japan Society for the Promotion of Science (JSPS) 
KAKENHI Grant-in-Aid for Scientific Research on Innovative Areas 25109004 (T.G.T. \& T.T.), 25109003 (S.K.) and 25109002 (Y.A.),  
Grant-in-Aid for Scientific Research (A) 15H02090 (T.G.T.), 
Challenging Exploratory Research 26610047 (T.G.T.), 
Grant-in-Aid for Young Scientists (B) 15K17648 (A.T.) and Grant-in-Aid for JSPS Fellows 15J01842 (H.M.). 
This study was also supported by the VLSI Design and Education Center (VDEC), the University of Tokyo in collaboration with Cadence Design Systems, Inc., and Mentor Graphics, Inc.


\end{document}